\documentclass[twocolumn,prb,aps,showpacs,amsmath,amssymb,floatfix]{revtex4}
\usepackage{bm}
\usepackage{graphicx}

\begin{document}

\title{Generalized joint density of states and its application to exploring the pairing symmetry of 
superconductors}

\author{Dan-Bo Zhang}
\affiliation{Department of Physics, Renmin University of China, Beijing, China} %

\author{Qiang Han}
\affiliation{Department of Physics, Renmin University of China, Beijing, China} %
\affiliation{Department of Physics and Center of Theoretical and Computational Physics, The University of Hong Kong, Pokfulam Road, Hong Kong China}%

\author{Z.~D.~Wang}
\affiliation{Department of Physics and Center of Theoretical and Computational Physics, The University of Hong Kong, Pokfulam Road, Hong Kong China}%

\begin{abstract}

We introduce a generalized joint density of states (GJDOS), which incorporates the coherent factor into the JDOS, to study quasiparticle interference (QPI) in superconductors. The intimate relation between the Fourier-transformed local density of states and GJDOS is revealed: they correspond respectively to the real and imaginary parts of a generalized impurity-response function, and particularly share the same angular factors and singular boundaries, as seen from our approximate analytic results for $d$-wave superconductors.
Remarkably, our numerical GJDOS analysis agrees well with the QPI patten of $d$-wave cuprates and $s_\pm$-wave iron-based superconductors.
Moreover, we illustrate that the present GJDOS scenario can uncover the sign features of the superconducting gap and thus can be used to explore the possible pairing symmetry of the K$_x$Fe$_{2-y}$Se$_2$
superconductors.

\end{abstract}
\maketitle

In superconductors, scattering of Bogoliubov quasiparticles by impurities can form quasiparticle interference (QPI) patterns which can be examined by the Fourier transformed scanning tunneling spectroscopy (FT-STS)\cite{Hoffman,McElroy03nature}. For high-$T_c$
cuprates the characteristic wave vectors as well as their dispersions
measured by the FT-STS experiments are interpreted quite well by the octet model \cite{McElroy03nature,qhwang}, which is based on the joint density of states (JDOS) argument. However, JDOS alone is incapable of explaining the variation of the QPI patterns when the magnetic field is applied. Indeed it was observed that QPI peaks corresponding to certain octet wave vectors are suppressed or enhanced by the magnetic field depending on whether the sign of the SC gaps is reversed or preserved~\cite{T. Hanaguri1}. The reason lies in that the probability of the elastic scattering of Bogoliubov quasiparticles on the constant contour of energy (CCE) is governed not only by their JDOS but also by the matrix element of the impurity Hamiltonian. Considering that  Bogoliubov quasiparticles in superconductors carry proper coherent factors reflecting their particle-hole dualism, this scattering matrix element will bear a particular combination of the coherent factors relying on the type of the scatterers\cite{franz03}. Taking into account an important role of coherent factors, the sign-sensitive effect of the QPI pattern on the magnetic field is naturally explained in the sense that the Andreev scattering of quasiparticles is enhanced by the pair-breaking impurities (mainly vortices) induced by the field. Notably, this sign-sensitive effect has also been applied to verify the $s_{\pm}$ pairing symmetry for iron-based superconducting (SC) material Fe(Se,Te)\cite{T. Hanaguri2}.

The insufficiency of the JDOS argument motivates us to propose the generalized JDOS (GJDOS) in this paper, which intrinsically incorporates the coherence factors.
We find that the GJDOS and the Fourier-transformed local density of states (FT-LDOS) can be derived from the same quantity.  The close connection between GJDOS and the Fourier-transformed local density of states (FT-LDOS) are established, which confirmed by our analytic results of $d$-wave superconductors under the Dirac cone approximation. Our numerical results of GJDOS capture the essential structure of the QPI pattern and explain its evolution with the magnetic field for both $d$-wave cuprates and $s_\pm$-wave iron-based superconductors. Finally, to explore the pairing symmetry of K$_{x}$Fe$_{2-y}$Se$_{2}$, we evaluate GJDOS numerically and find that the results are distinct for the theoretically-proposed nodeless $d$-, nodal extended $s$- and $s_{++}$-wave SC states. Thus the pairing symmetry of K$_{x}$Fe$_{2-y}$Se$_{2}$ may be identified by checking the QPI pattern under the magnetic field via the FT-STS experiments.

\label{relation between GJDOS and FT-LDOS}
 We first develop the basic formalism for both the FT-LDOS and GJDOS.
For different types of weak impurity scatterers, the FT-LDOS can be divided into the corresponding channels under the Born approximation~\cite{scalapino},
\begin{equation}
\label{FT-LDOS}
     S_i(\mathbf{p},\omega) = -\frac{1}{\pi}\text{Im} [\sum_\mathbf{k}G(\mathbf{k+p},\omega^+)\sigma_i G(\mathbf{k},\omega^+ ]_{11}
\end{equation}
where $\omega^+\equiv\omega+i0^+$. $\sigma_i$ are Pauli matrices in Nambu space and correspond to magnetic ($i=0$), pair-breaking ($i=1,2$)~\cite{franz03} and nonmagnetic ($i=3$) impurity potentials.  $G(\mathbf{k},z)=(z-\xi_\mathbf{k}\sigma_3-\Delta_\mathbf{k}\sigma_1)^{-1}$ is the Nambu-Gor'kov Green's function. Here the SC gap $\Delta_\mathbf{k}$ is a real function for high-$T_c$ superconductors, and therefore $S_2$ is zero according to Eq.~(\ref{FT-LDOS}) for the case of $i=2$.
Thus, we only discuss the $i=0,1,3$ cases hereafter.

In order to deliver the concept of GJDOS and see how the FT-LDOS is related to GJDOS, we introduce a generalized impurity response function $L_i$
\begin{equation}
    L_i(\mathbf{p},\omega) = \sum_{s=\pm,\mathbf{k}}G_{1\alpha}(\mathbf{k}+s\mathbf{p},\omega^+) \sigma _i^{\alpha \beta } A_{\beta}(\mathbf{k},\omega)
    \label{define_L}
\end{equation}
where $A_\alpha(\mathbf{k},\omega)=-\text{Im}G_{\alpha 1}(\mathbf{k},\omega^+)/\pi$ with $A_1$ and $A_2$ correspond to the imaginary part of the normal and anomalous Green's functions, respectively. From $L_i$ we can derive two quantities,
\begin{equation}
\label{define of S and J}
\begin{aligned}
S_i(\mathbf{p},\omega)=\text{Re} L_i(\mathbf{p},\omega ) \\ J_i(\mathbf{p},\omega)=-\frac{1}{\pi}\text{Im} L_i(\mathbf{p},\omega ).
\end{aligned}
\end{equation}
where $S_i(\mathbf{p},\omega)$ is identical to Eq.~(\ref{FT-LDOS}), while $J_i(\mathbf{p},\omega)$ represents our GJDOS. It is clearly shown in Eq.~(\ref{define of S and J}) that the intrinsic relation between FT-LDOS and GJDOS lies in that they are just the real and imagine part of the quantity $L_i(\mathbf{p},\omega^+ )$ in analogy to, e.g., the dielectric response function.
In addition to the autocorrelation of the normal spectral function $A_1$\cite{McElroy}, GJDOS also involves the autocorrelation of the anomalous spectral function $A_2$ as well as the correlation between $A_1$ and $A_2$. Therefore GJDOS extends the usual JDOS used in previous works\cite{McElroy03nature,McElroy,franzl,han,note}, and is more transparent than the complicated $J(\mathbf{q},E,E^\prime)$ used in Ref.[\onlinecite{Marianna Maltseva}].

Now we show how the coherent factors are naturally and appropriately embodied in GJDOS. From
 $A_1(\mathbf{k},\omega) = u_\mathbf{k}^2\delta(\omega-E_\mathbf{k}) + v_\mathbf{k}^2\delta(\omega+E_\mathbf{k})$ and $A_2(\mathbf{k},\omega)=u_\mathbf{k}v_\mathbf{k}[\delta(\omega-E_\mathbf{k})-\delta(\omega+E_\mathbf{k})]$,
 where
  $u_\mathbf{k}^2= (1+\xi_\mathbf{k}/E_\mathbf{k})/2$,
  $v_\mathbf{k}^2 =(1-\xi_\mathbf{k}/E_\mathbf{k})/2$,
  $u_\mathbf{k}v_\mathbf{k}=\Delta_\mathbf{k}/2E_\mathbf{k}$,
  $E_\mathbf{k} = \sqrt{\xi_{\mathbf{k}}^2+\Delta_\mathbf{k}^2}$, and according to Eq.~(\ref{define_L}) and (\ref{define of S and J}),
 we obtain  for $\omega>0$ \cite{note1},
\begin{equation}
\label{GJDOS}
\begin{aligned}
 J_i(\mathbf{p},\omega) =2\sum_{\mathbf{k}}C_i(\mathbf{k}+\mathbf{p},\mathbf{k})\delta(\omega-E_{\mathbf{k}+\mathbf{p}})\delta(\omega-E_\mathbf{k})\\
\end{aligned}
\end{equation}
where $C_i(\mathbf{k'},\mathbf{k})$ is the coherent factor as given in Table \ref{table of coherent factor} for three types of impurity.
\begin{table}[htb]
\caption{Coherent factors for three types of impurity. }
\begin{tabular}{|c|c|c|c|}
  \hline
  impurity type   &coherent factor & enhanced & suppressed\\
  \hline
  \begin{tabular}{c}nonmagnetic \\  $i=3$ \end{tabular}   & $ u_\mathbf{k'}u_\mathbf{k}(u_\mathbf{k'}u_\mathbf{k}-v_\mathbf{k'}v_\mathbf{k})$ & $+-$    &  $++$ \\\hline
  \begin{tabular}{c}magnetic \\  $i=0$  \end{tabular}  & $ u_\mathbf{k'}u_\mathbf{k}(u_\mathbf{k'}u_\mathbf{k}+v_\mathbf{k'}v_\mathbf{k})$ & $++$ &$+-$\\\hline
  \begin{tabular}{c}pair-breaking \\  $i=1$ \end{tabular} & $ u_\mathbf{k'}u_\mathbf{k}(u_\mathbf{k'}v_\mathbf{k}+u_\mathbf{k}v_\mathbf{k'})$ &  $++$& $+-$\\\hline
\end{tabular}
\label{table of coherent factor}
\end{table}
Also shown are the sign sensitivities of the coherent factors for different quasiparticle scattering processes from $\mathbf{k}$ to $\mathbf{k}'$, where the sign of the superconducting gap is preserved or reversed indicated by $++$ ($--$) or $+-$ ($-+$). For the pair-breaking impurity case, the coherent factor is proportional to $\Delta_{\mathbf{k}}+\Delta_{\mathbf{k'}}$ for $\xi_\mathbf{k}\approx \xi_{\mathbf{k}^\prime}$, which leads to a striking effect that the scattering amplitude will be zero if $\Delta_{\mathbf{k}}$ and $\Delta_{\mathbf{k'}}$ having the same magnitude but opposite sign. This sign sensitivity of the coherent factor for different type of impurity has been observed in the vortex region of $d$-wave cuprates~\cite{T. Hanaguri1} and has also been applied to verify~\cite{T. Hanaguri2} the $s_\pm$-wave pairing symmetry of iron-based superconductor Fe(Se,Te).

To further investigate the relation between FT-LDOS and GJDOS, we turn to derive some analytic results of the complex response function $L_i(\mathbf{p},z)$ for $d$-wave superconductors under the Dirac cone approximation.  Around the node we make the linear expansions with $\xi_\mathbf{k}\approx v_\text{F} k_1$, $\Delta_\mathbf{k}\approx -v_\Delta k_2$,
and $E_\mathbf{k}\approx \sqrt{(v_\text{F} k_1)^2+(v_\Delta k_2)^2}$. For the intra-node scattering, we obtain from Eq.~(\ref{define_L}),
\begin{equation}
L_i(\mathbf{p},\omega)=-\frac{1}{4\pi v_\Delta v_\text{F}} \left\{
\begin{array}{ll}
z^2 (1-z^2)^{-\frac{1}{2}} \cos^2\theta, & i=3 \\
1-\sqrt{1-z^2}, & i=0 \\
z^2 (1-z^2)^{-\frac{1}{2}} \frac{\sin(2\theta)}{2}, & i=1
\end{array},
\right.
\nonumber
\end{equation}
where $\theta$ devotes the angle between $\mathbf{p}$ and the $[1\bar{1}0]$ direction, and $z=2\omega/E_\mathbf{p}$. From the above equation and Eq.~(\ref{define of S and J}), we can see that FT-LDOS and GJDOS, although not proportional to each other, share the same angular factors and singular boundaries, which holds true even in the inter-node scattering case. These common features can explain why GJDOS is appropriate to explain QPI in superconductors.
We find that our results of $L_i(\mathbf{p},\omega)$ agree with those obtained from $\Lambda(\mathbf{p},\omega)$ in Ref.~[\onlinecite{franz03,franzl}] for the nonmagnetic and magnetic cases.

We also find that our analytic result for intra-node scattering in $d$-wave superconductors by magnetic impurities is very similar to that for scattering in surface of 3D topological insulators~\cite{guo} and intra-valley scattering in graphene~\cite{T. Pereg-Barnea} by nonmagnetic impurities.
The physics lies in the common Dirac cone structure and the $\sigma_0$ scattering channel.
For illustration, we express for $d$-wave superconductors in Nambu space $\hat{H}_\mathbf{k}=\xi_\mathbf{k}\sigma_3+\Delta_\mathbf{k}\sigma_1 \approx v_\text{F} k_1 \sigma_3 - v_\Delta k_2 \sigma_1 = -[(v_\text{F}k_1,0,v_\Delta k_2)\times \bm{\sigma}]\cdot \mathbf{n}_2 $, which is equivalent locally to an anisotropic Rashba spin-orbit-coupling Hamiltonian for the surface states of topological insulators.

\begin{figure}[ht]
\begin{tabular}{cc}
(a) &  (b) \\
\includegraphics[width=0.20\textwidth]{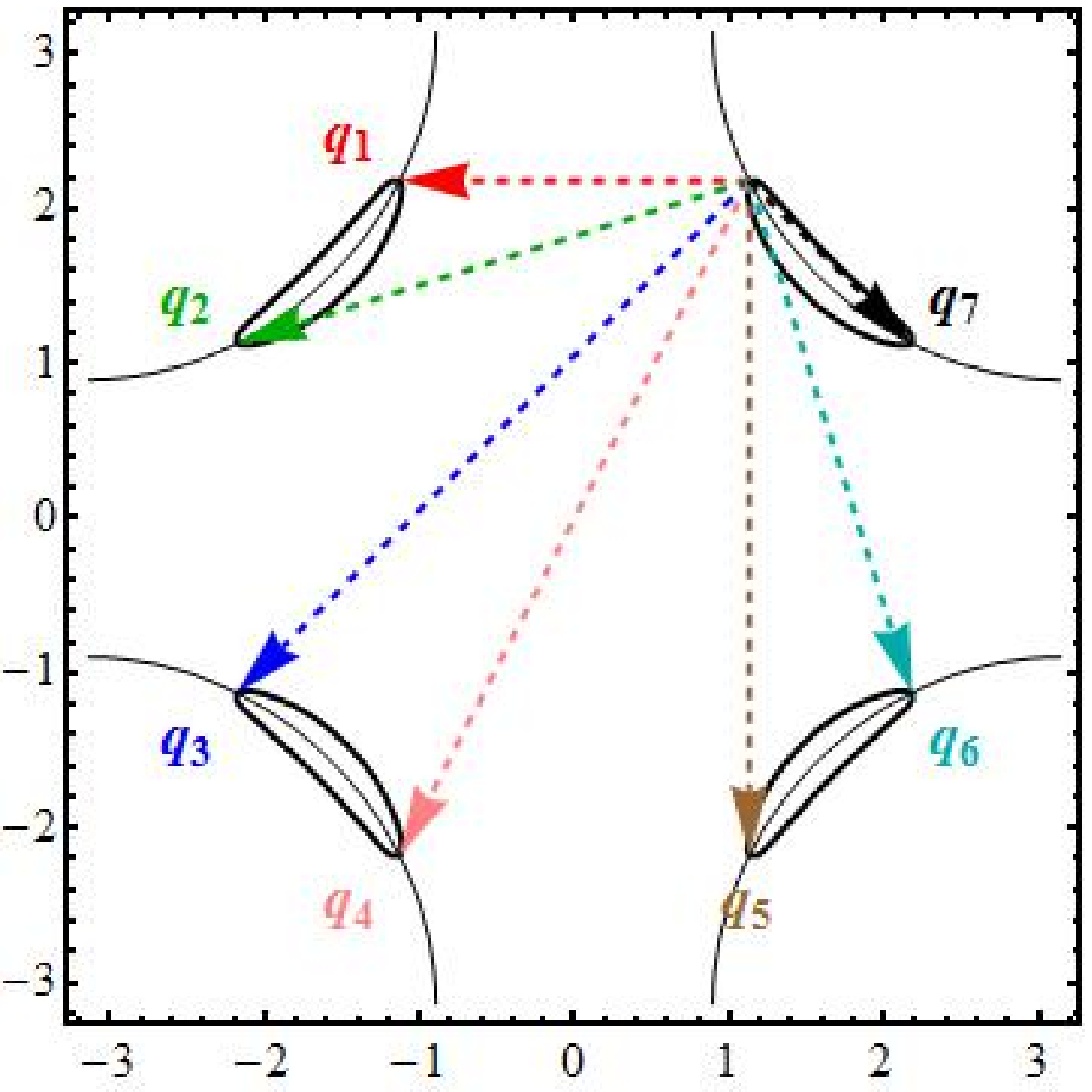} & \includegraphics[width=0.20\textwidth]{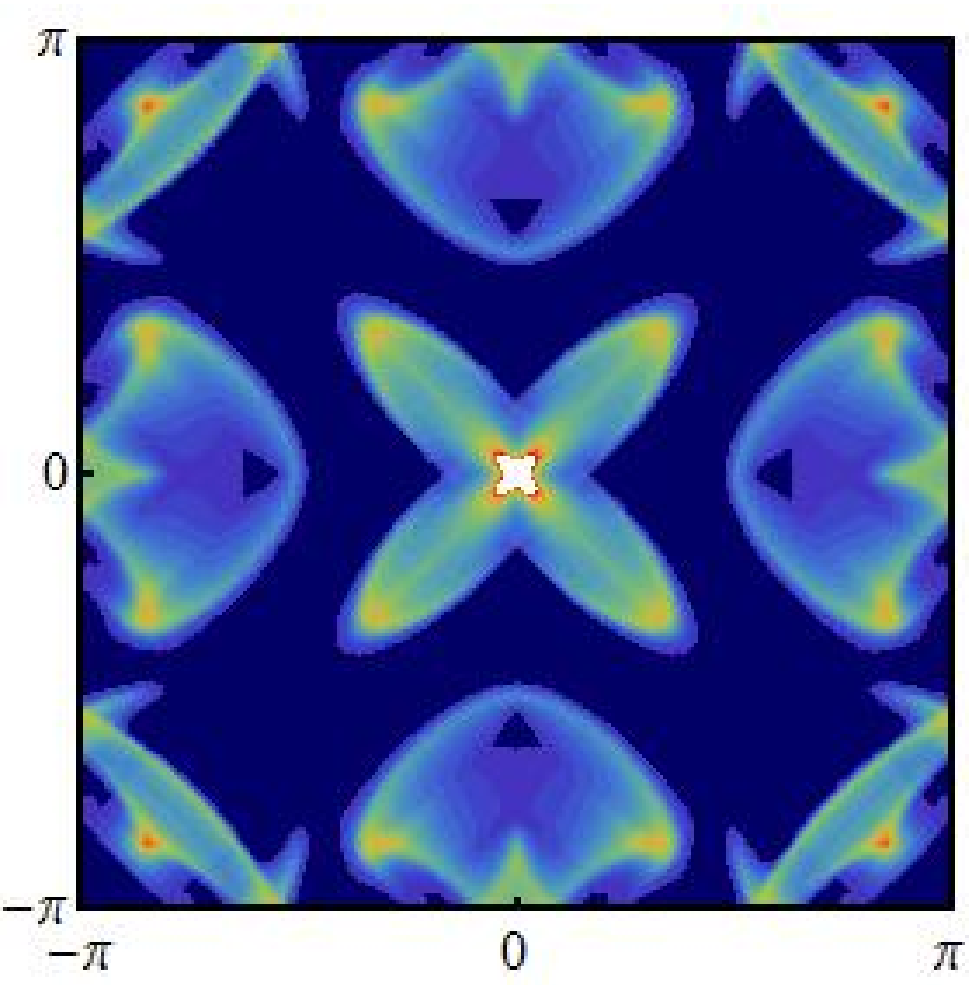} \\
(c) & (d) \\
\includegraphics[width=0.20\textwidth]{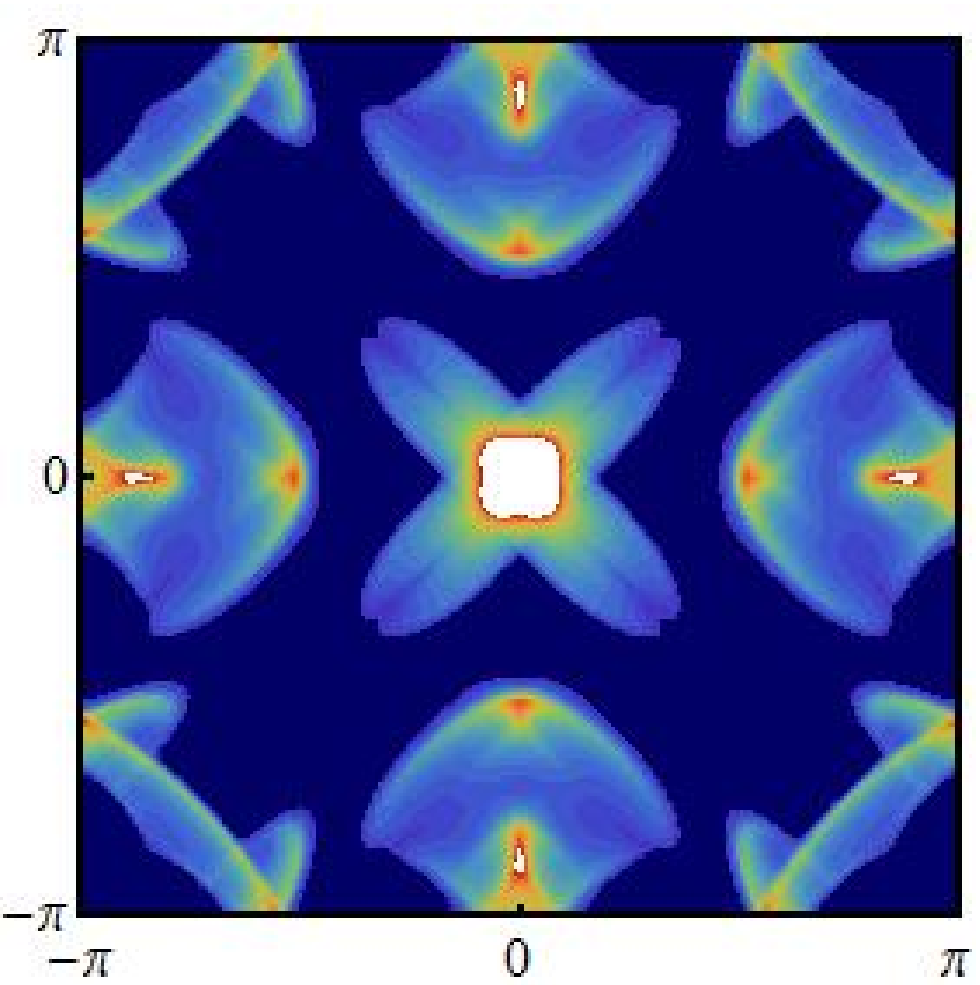} & \includegraphics[width=0.20\textwidth]{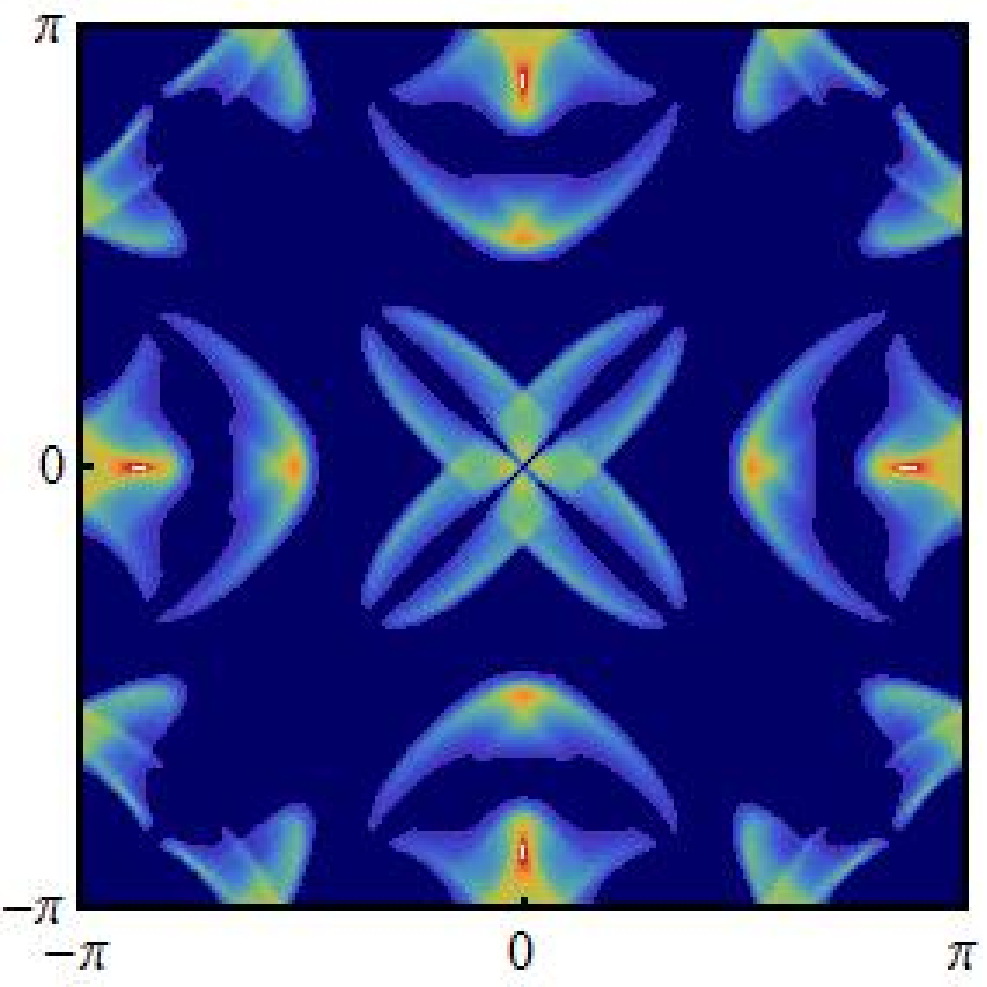} \\
\end{tabular}
\includegraphics[width=0.25\textwidth]{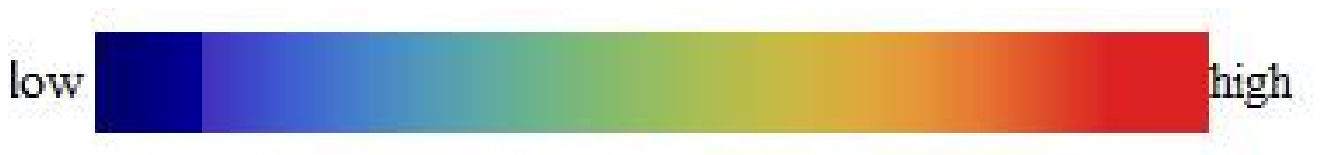}
\caption{ (Color online) (a) CCE in $d$-wave superconductors under Dirac cone approximation.
The GJDOS patterns  calculated according to Eq.~(\ref{GJDOS}) for (b) nonmagnetic (c) magnetic and (d) pair-breaking ($i=1$) impurity cases. The normal-state dispersion is $\xi_\mathbf{k}=-2t(\cos k_x+\cos k_y)-4t^\prime\cos k_x \cos k_y - \mu$ and the gap function $\Delta_\mathbf{k}=2\Delta_0(\cos k_y - \cos k_x)$ with parameters $ t=1,t^\prime=-0.3,\mu=-1,\Delta_0=0.2,\omega=0.4$.
$\delta(x) \approx \frac{\eta_0}{\pi(x^2+\eta_0^2)}$ with $\eta_0=0.01$ the broadening parameter.  }
\label{fig1}
\end{figure}

The GJDOS for $d$-wave superconductors are calculated numerically according to Eq.~(\ref{GJDOS}) and the results are shown in Fig.~\ref{fig1}. $\mathbf{q}_2$, $\mathbf{q}_3$, $\mathbf{q}_6$, $\mathbf{q}_7$ ($\mathbf{q}_1$, $\mathbf{q}_4$, $\mathbf{q}_5$) are relatively brighter (darker) spots  for the nonmagnetic case contrary to the magnetic case as shown in Fig.~\ref{fig1}(b) and (c), which is consistent to Table \ref{table of coherent factor}. For the pair-breaking impurity, the diminishing of the GJDOS peaks at positions $\mathbf{q}_2, \mathbf{q}_3, \mathbf{q}_6, \mathbf{q}_7$ and enhancement at positions $\mathbf{q}_1, \mathbf{q}_4, \mathbf{q}_5$ agree with the magnetic field dependence of the QPI pattern observed in high-$T_c$ cuprates~\cite{T. Hanaguri1}.

\begin{figure}[htb]
\begin{tabular}{cc}
(a) & (b) \\
\includegraphics[width=0.20\textwidth]{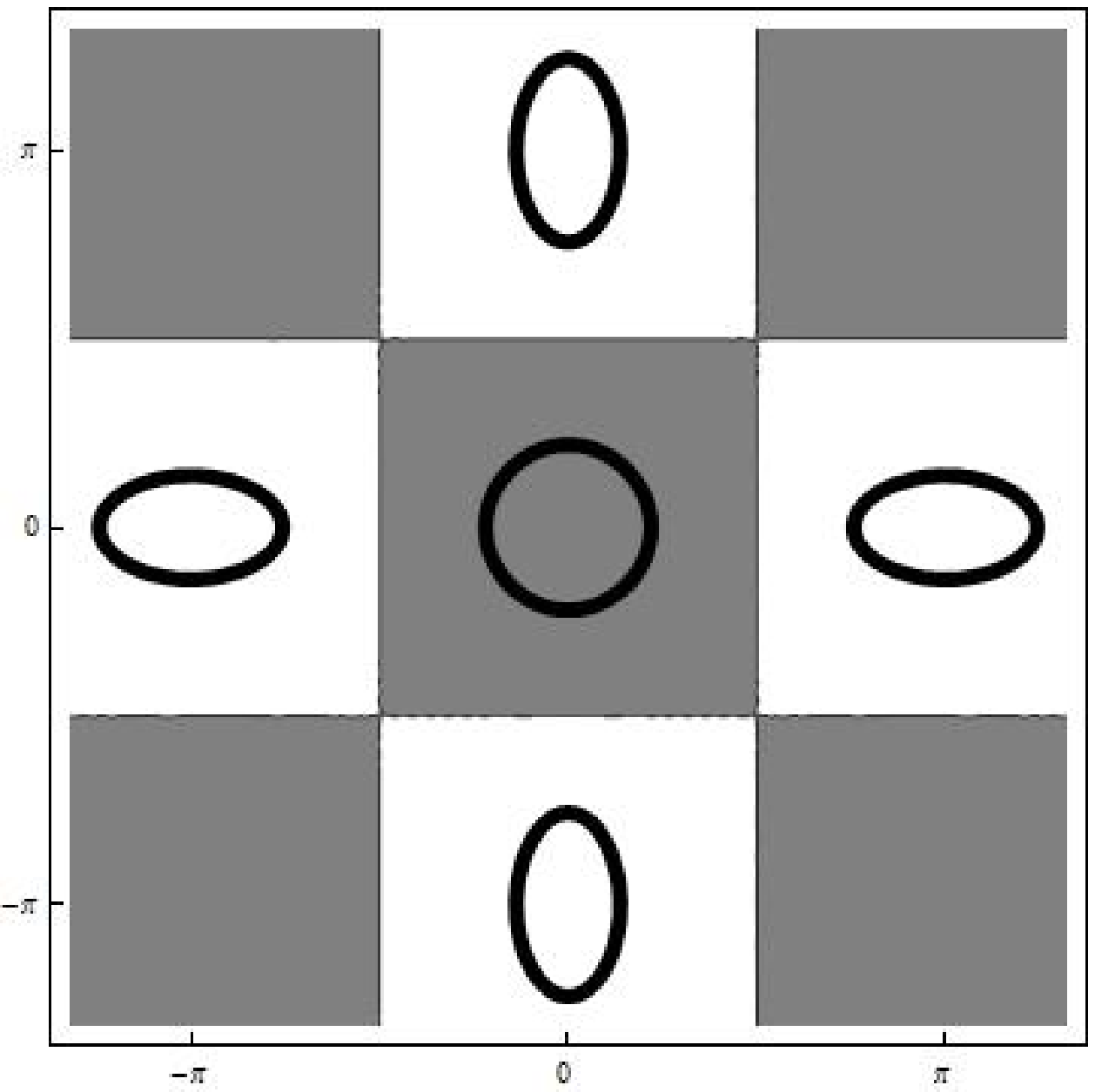} & \includegraphics[width=0.20\textwidth]{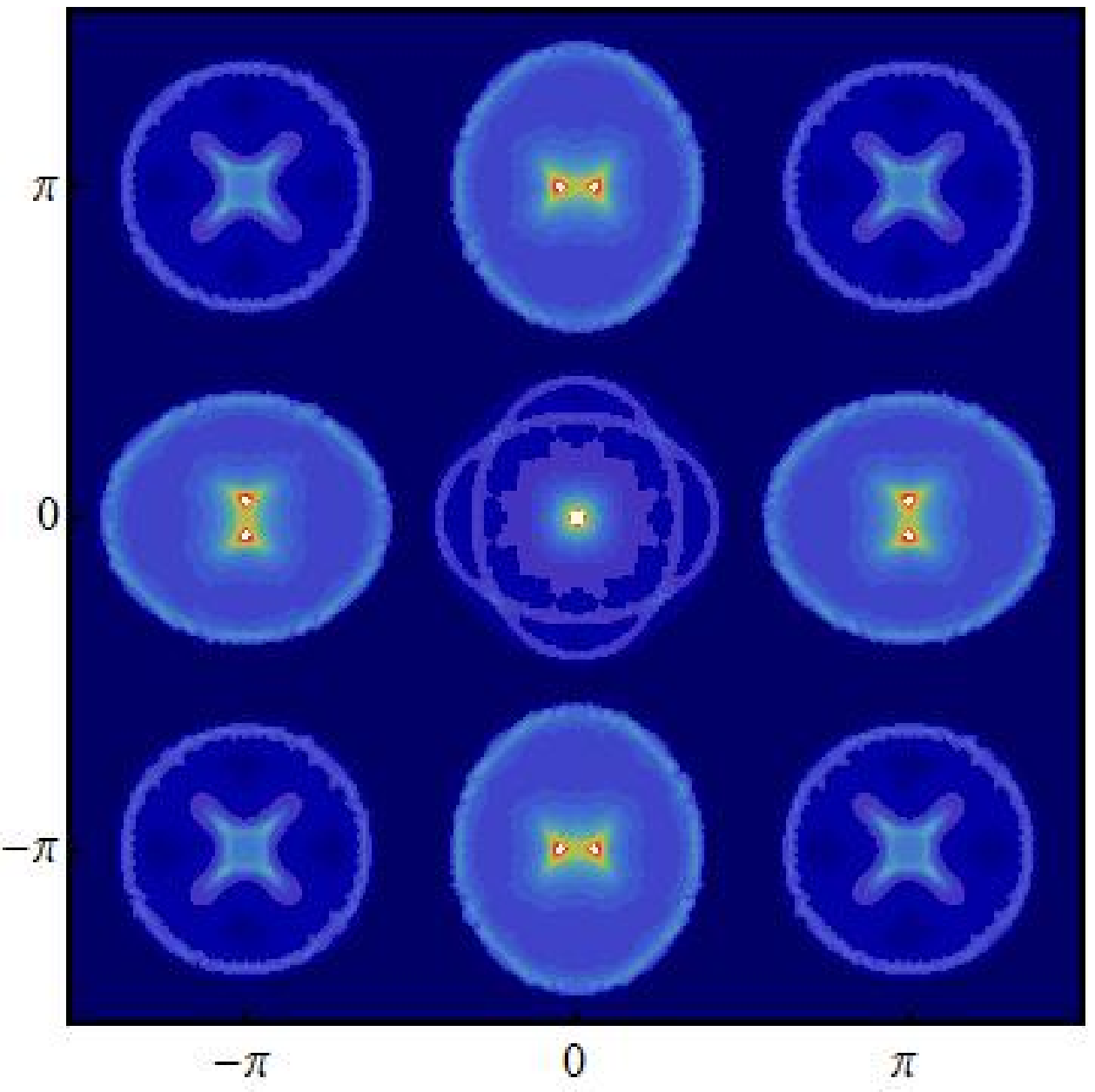}  \\
(c) & (d) \\
\includegraphics[width=0.20\textwidth]{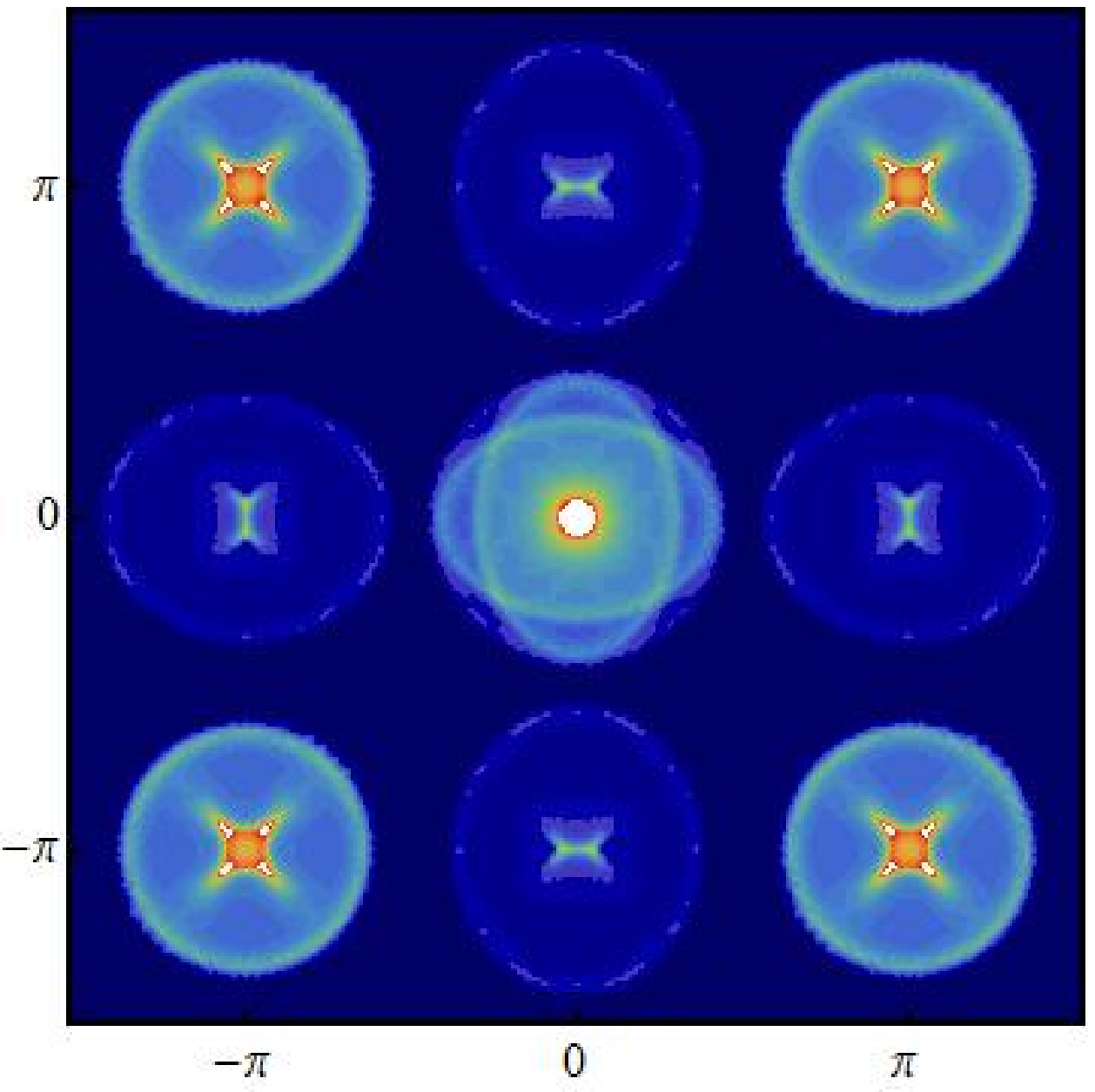} & \includegraphics[width=0.20\textwidth]{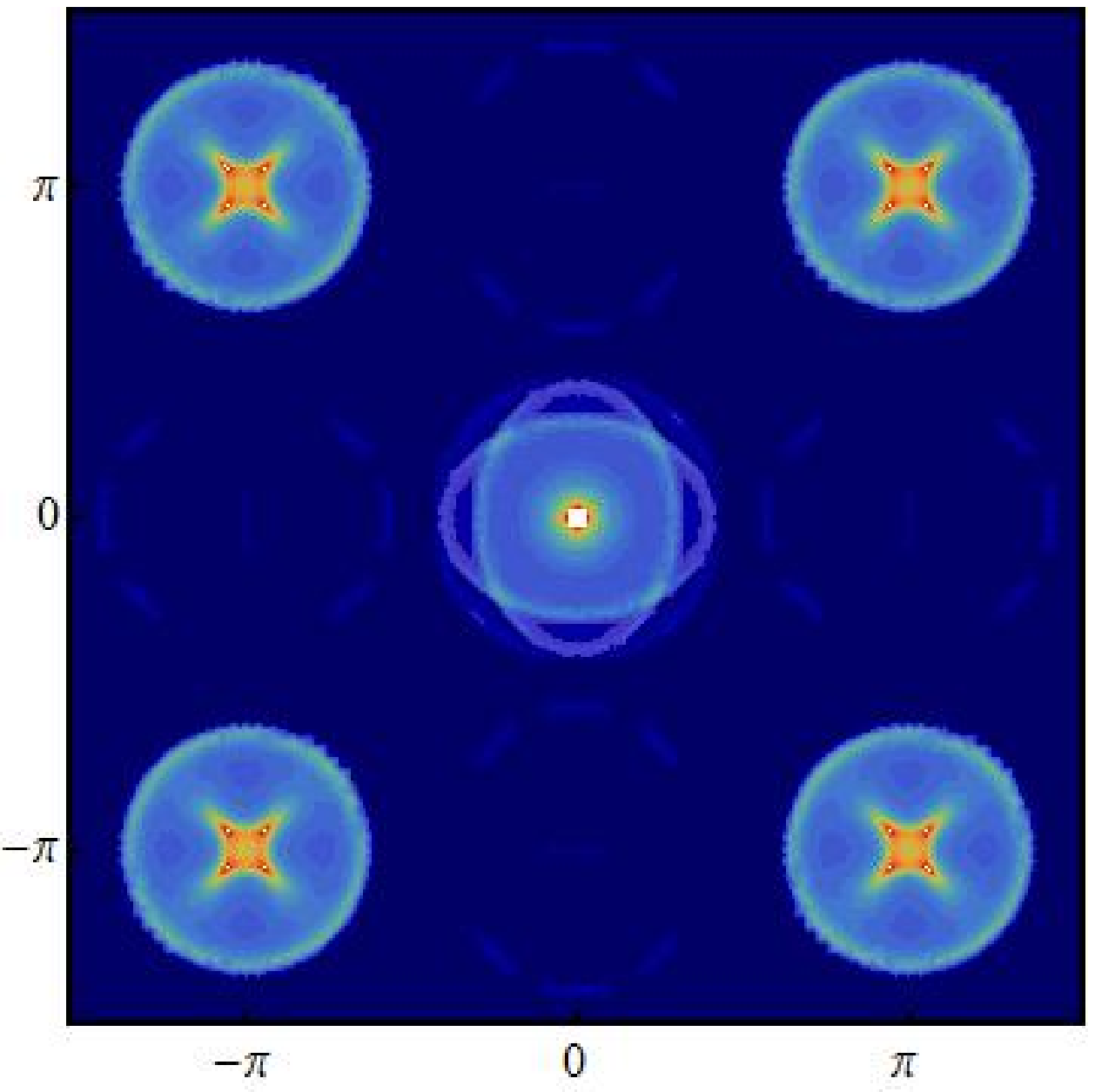}
\end{tabular}
\includegraphics[width=0.25\textwidth]{colorfunction.eps}
\caption{(Color online) (a) Fermi surface in Fe(Se,Te).  The gray/white color represents the positive/negative sign of the SC gap with $s_\pm$-wave symmetry. The GJDOS patterns for (b) nonmagnetic (c) magnetic and (d) pair-breaking ($i=1$) impurity cases. The normal-state dispersion \cite{A. Akbari} is $\xi_\mathbf{k}=\varepsilon_0-t_1[(1-\lambda)\cos{(k_x-\pi)}+(1+\lambda)\cos{k_y}]-\mu_1$
for $\mathbf{k}$ near $(\pi,0)$, $\xi_\mathbf{k}=\varepsilon_0-t_1[(1+\lambda)\cos{k_x}+(1-\lambda)\cos(k_y-\pi)]-\mu_1$
for $\mathbf{k}$ near $(0,\pi)$, and $\xi_\mathbf{k}=\varepsilon_0-t_2(\cos{k_x}+\cos{k_y})-\mu_2$ for $\mathbf{k}$ near $(0,0)$. The gap function is $\Delta \cos{k_x}\cos{k_y}$. The parameters are set as:
$\varepsilon_0=0.035, t_1=-0.68, t_2=0.85, \mu_1=-1.23, \mu_2=1.54, \lambda=0.3, \Delta=0.01, \omega=0.01$.}
\label{fig2}
\end{figure}
 We now turn to address QPI patterns in iron-based superconductor~\cite{T. Hanaguri2, hu jiangping, J. Knolle, Sykora, A. Akbari, Hess}. Let us first consider where QPI peaks are expected to appear. It should be noted that wave vectors that connect maximum density of state on CCEs as in $d$-wave cuprates are not expected here for the lack of strong anisotropy of CCEs for iron-based superconductors. However, nearly nesting of CCEs are typical for iron-based superconductor and QPI peaks would appear at those nesting vectors $\mathbf{q}$ where enormous scattering processes $\mathbf{k}\rightarrow \mathbf{k+q}$ on CCEs are involved. Those QPI peaks around nesting vectors are easily confused with peaks resulting from static orders such as the spin or charge density wave. However, QPI peaks are sensitive to coherent factors for different types of impurity depending on the relative gap sign for each pockets and can be suppressed or enhanced by introducing more pair-breaking scatterers after intentionally applying the magnetic field. This has been observed in the FT-STS experiment for Fe(Se,Te) that strong suppression of QPI peaks at $(\pm\pi,0)$ and $(0,\pm\pi)$ due to pair-breaking scattering occurs at the nesting vectors that connect the hole and electron pockets, supporting the $s_\pm$ paring symmetry proposed by the spin-fluctuation mechanism. In Fig.\ref{fig2} we show three QPI patterns for three types of impurity. And one can see that with increasing magnetic field and accordingly the strengthening of the pair-breaking scattering, the QPI pattern will change from Fig.\ref{fig2}(b) to Fig.\ref{fig2}(d), as observed by the experiment~\cite{T. Hanaguri1}.

\begin{figure*}[ht]
\begin{tabular}{cccc}
(a) & (b) & (c) & (d) \\
\includegraphics[width=0.18\textwidth]{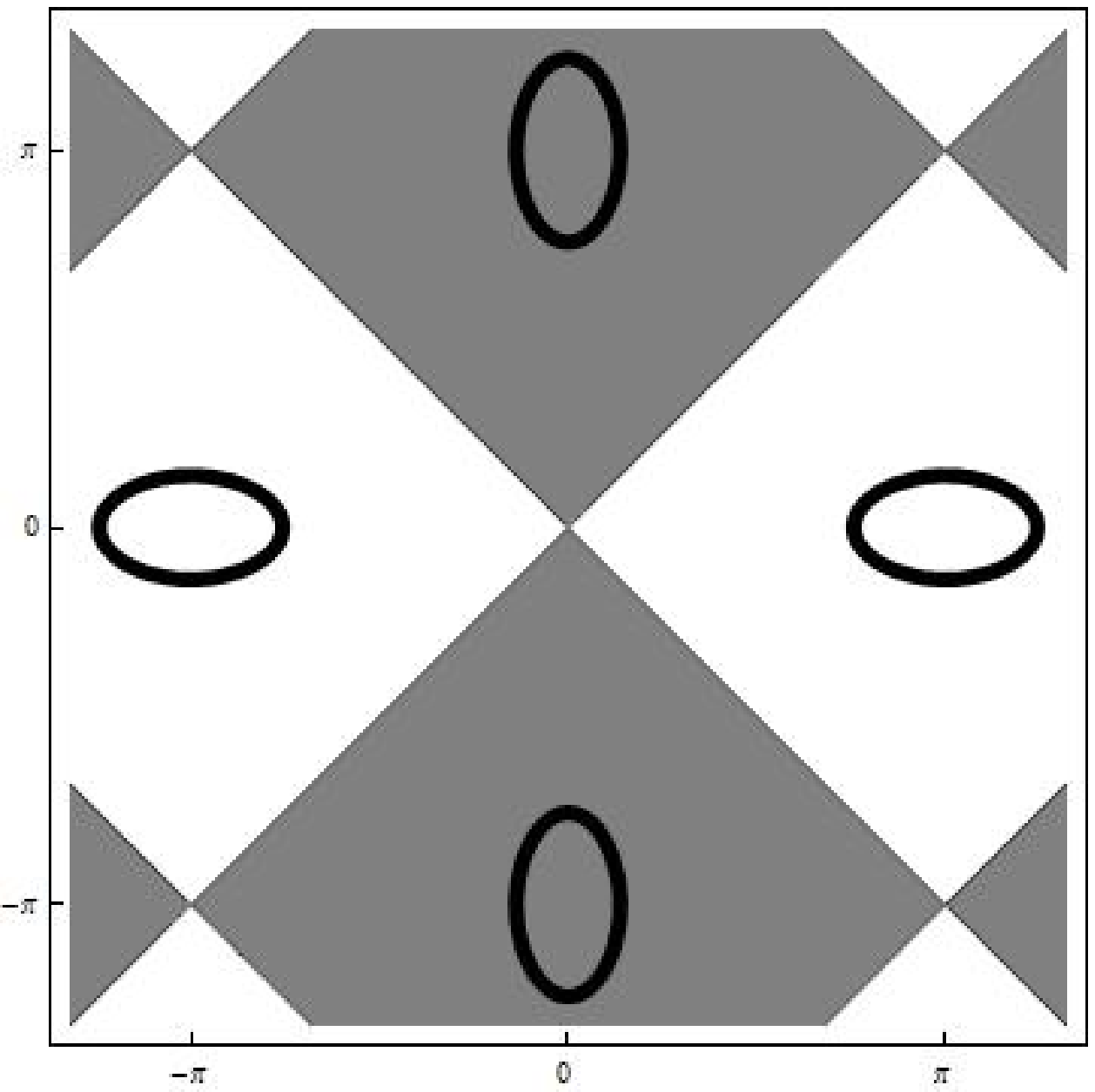} & \includegraphics[width=0.18\textwidth]{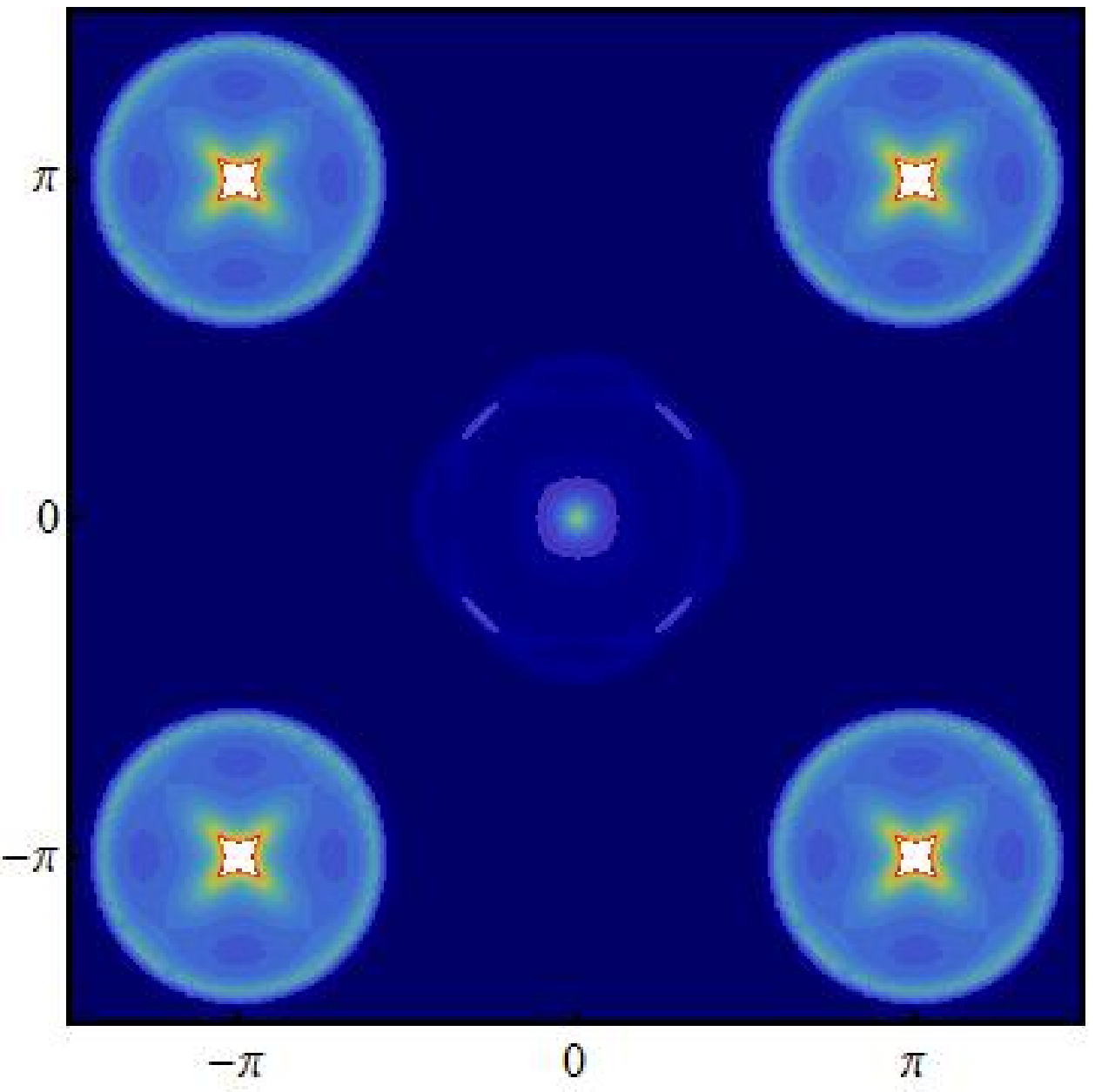} & \includegraphics[width=0.18\textwidth]{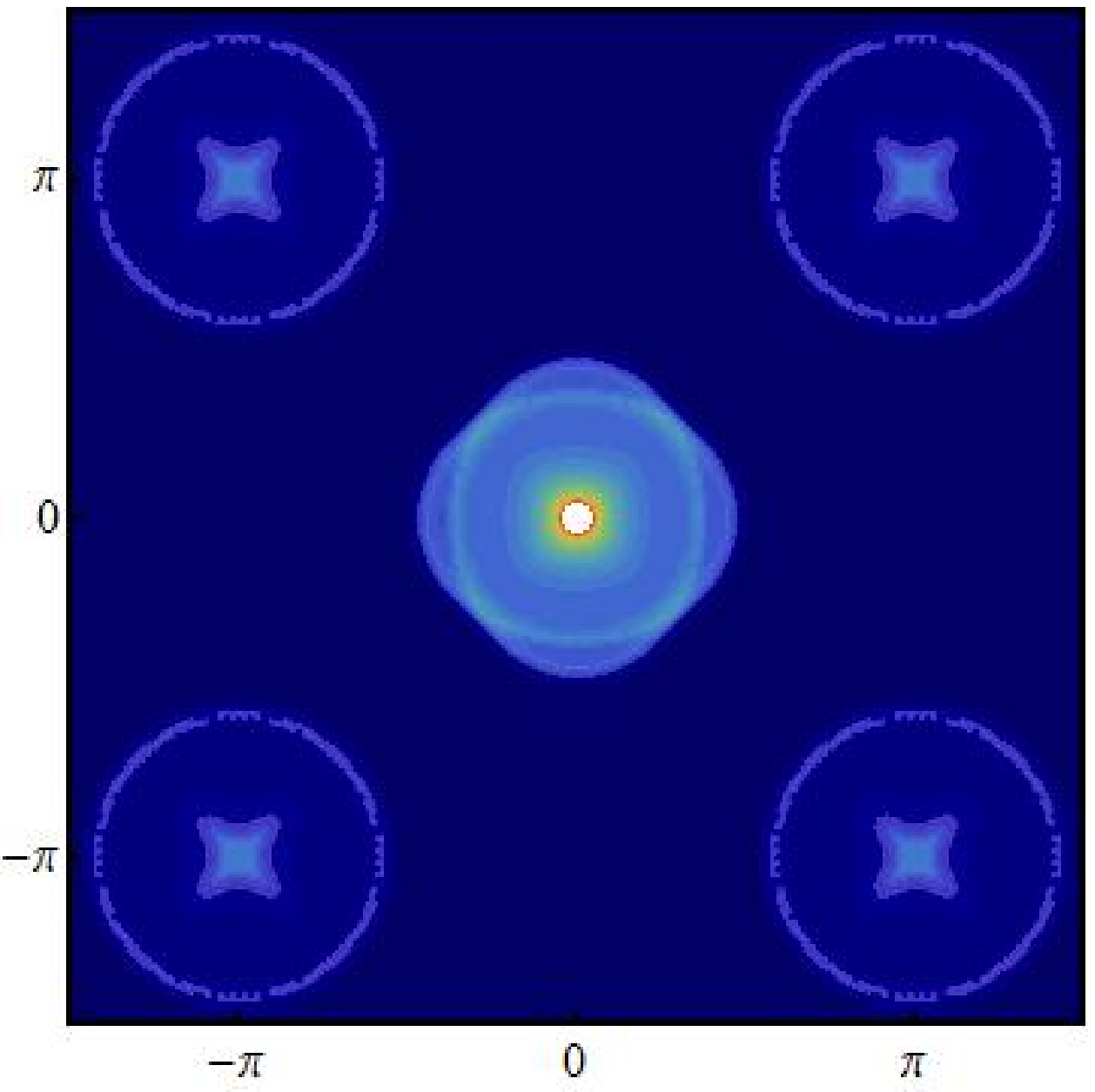} & \includegraphics[width=0.18\textwidth]{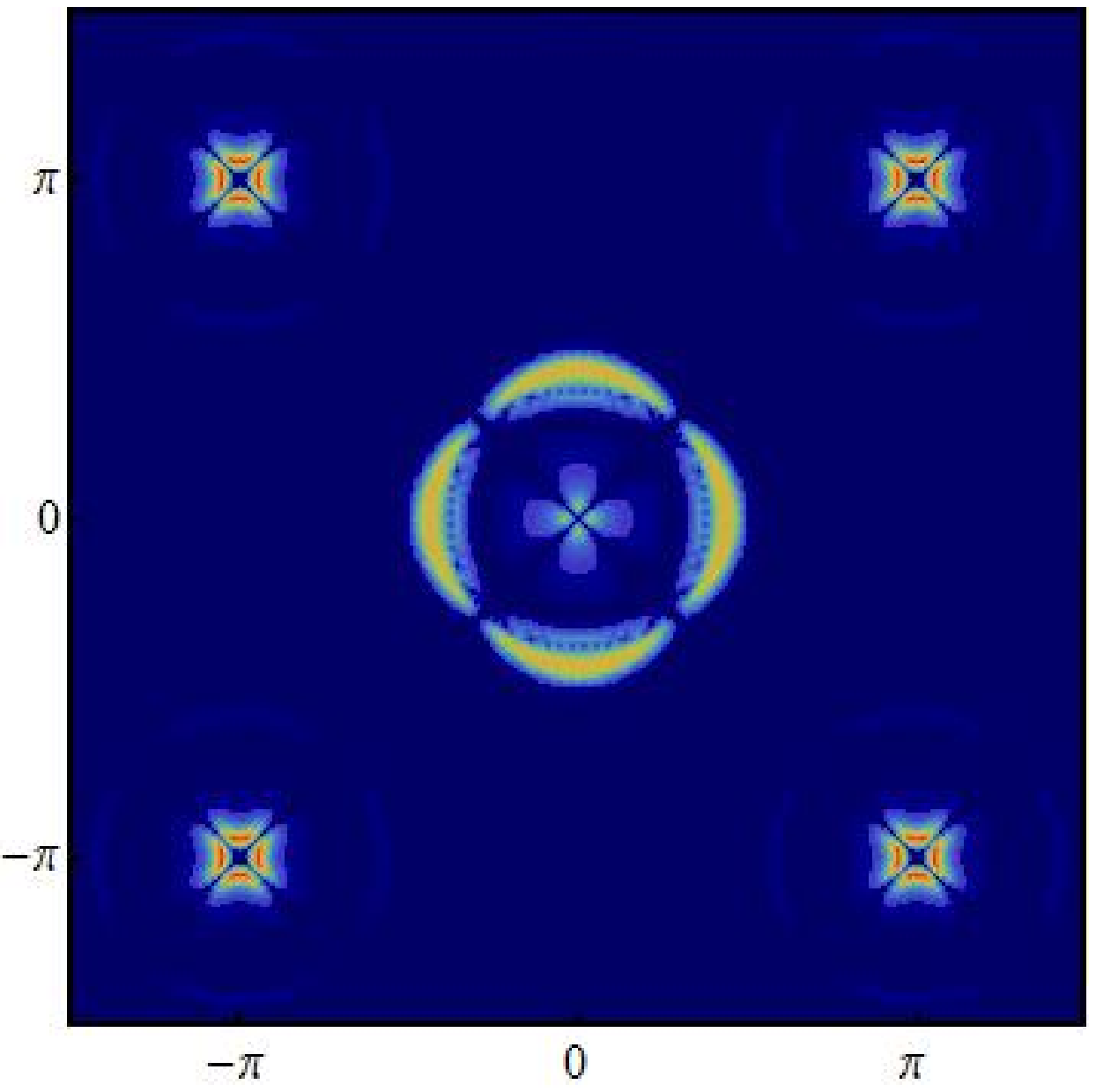} \\
\includegraphics[width=0.18\textwidth]{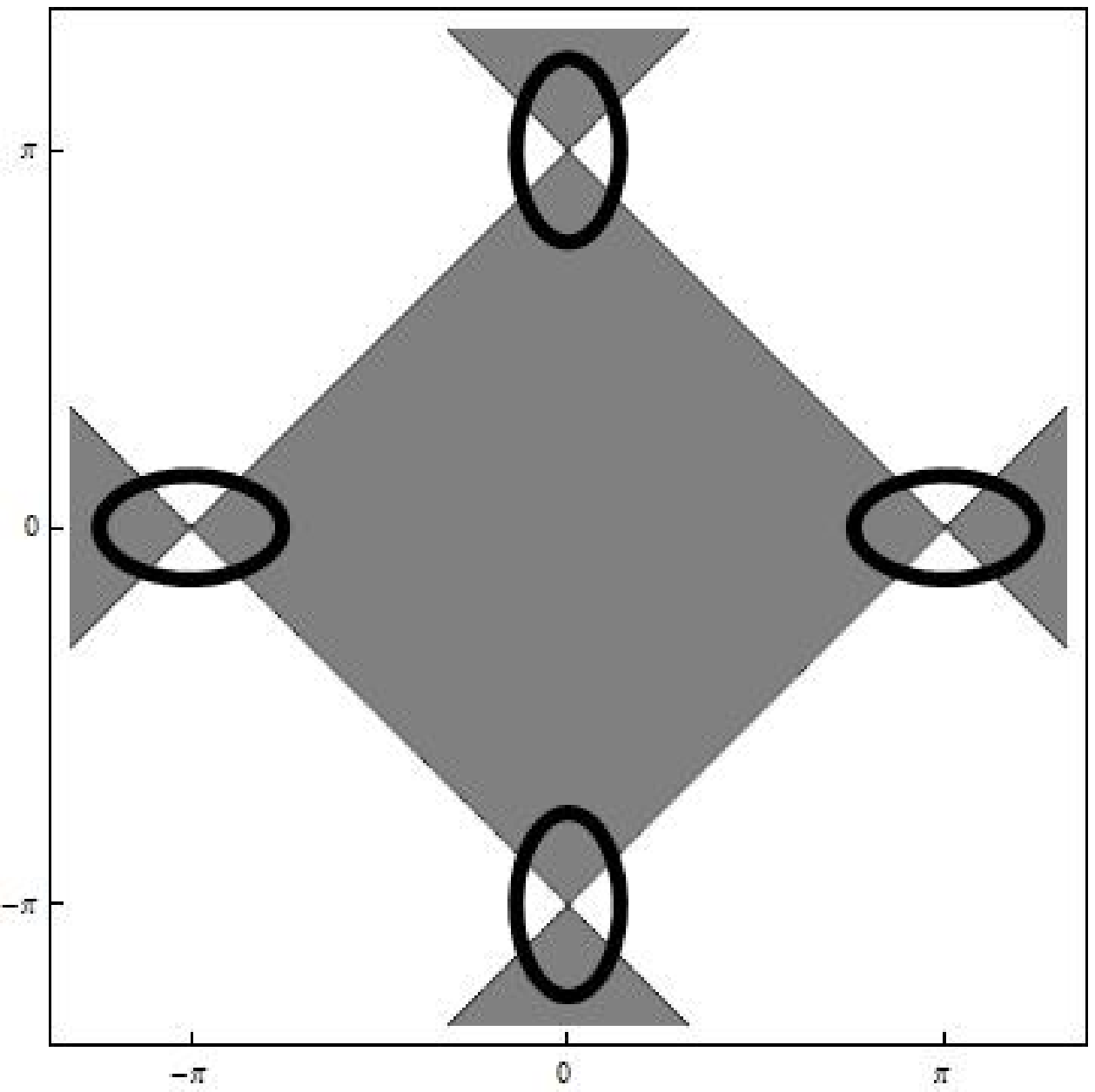} & \includegraphics[width=0.18\textwidth]{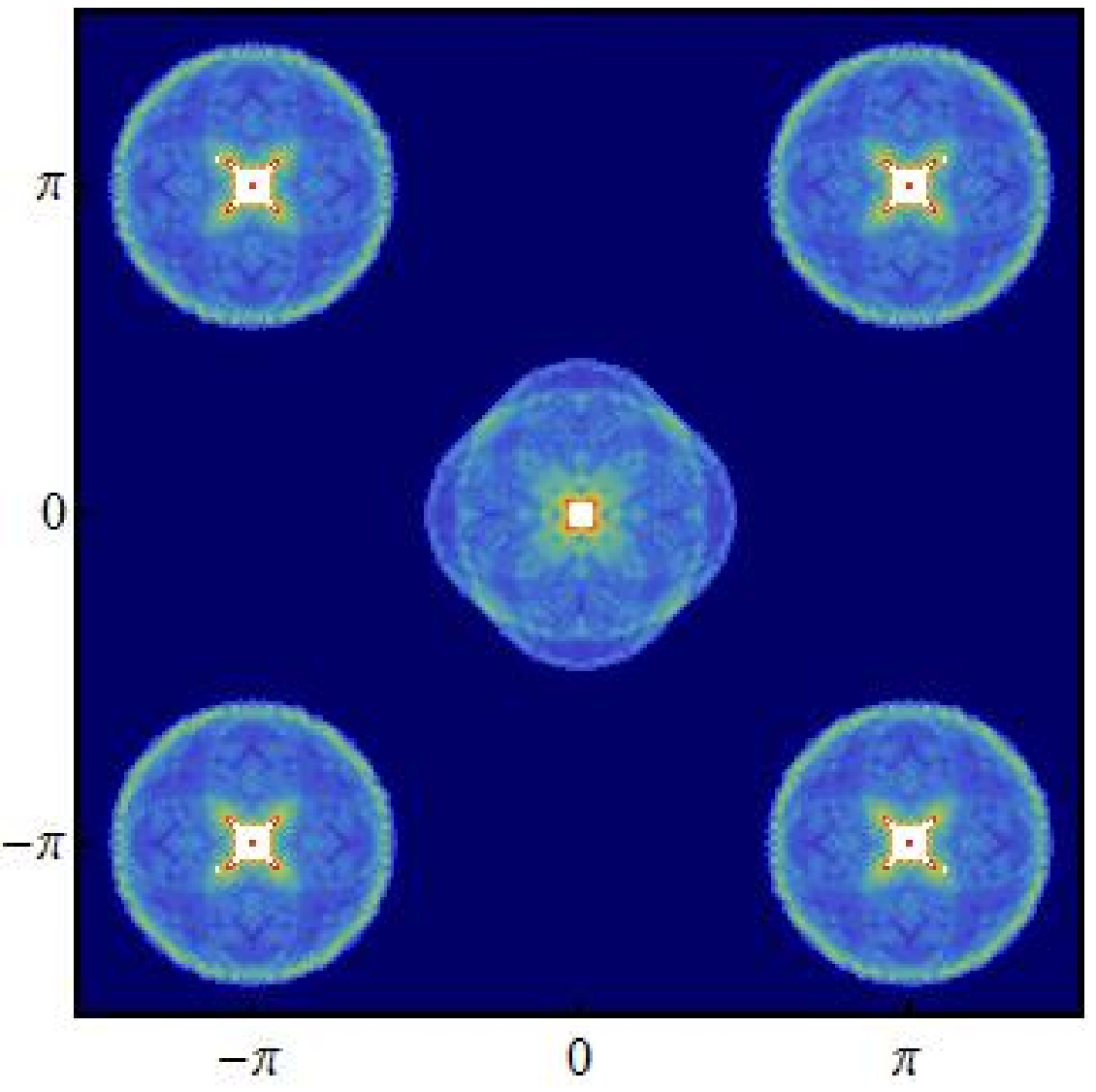} & \includegraphics[width=0.18\textwidth]{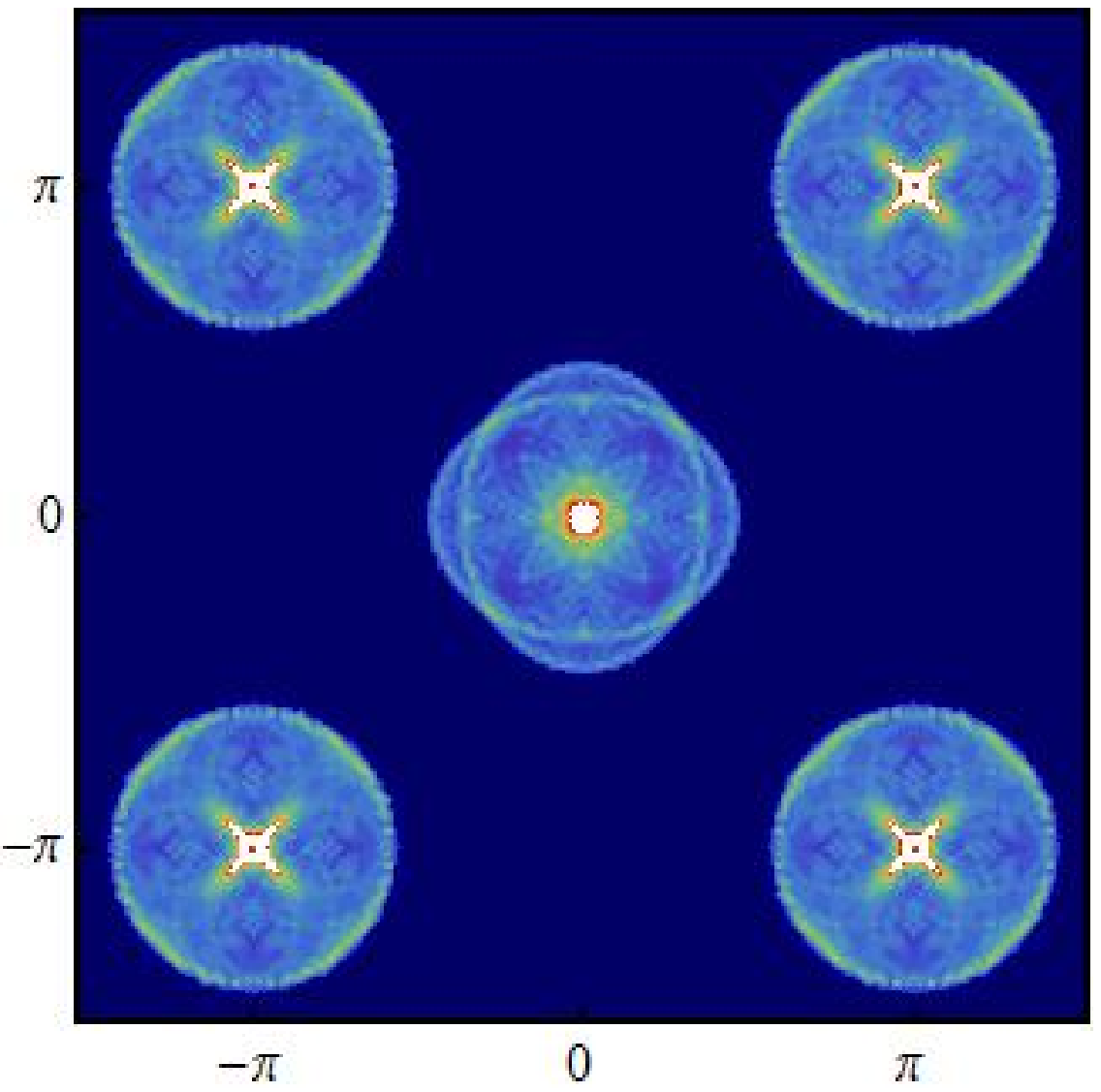} & \includegraphics[width=0.18\textwidth]{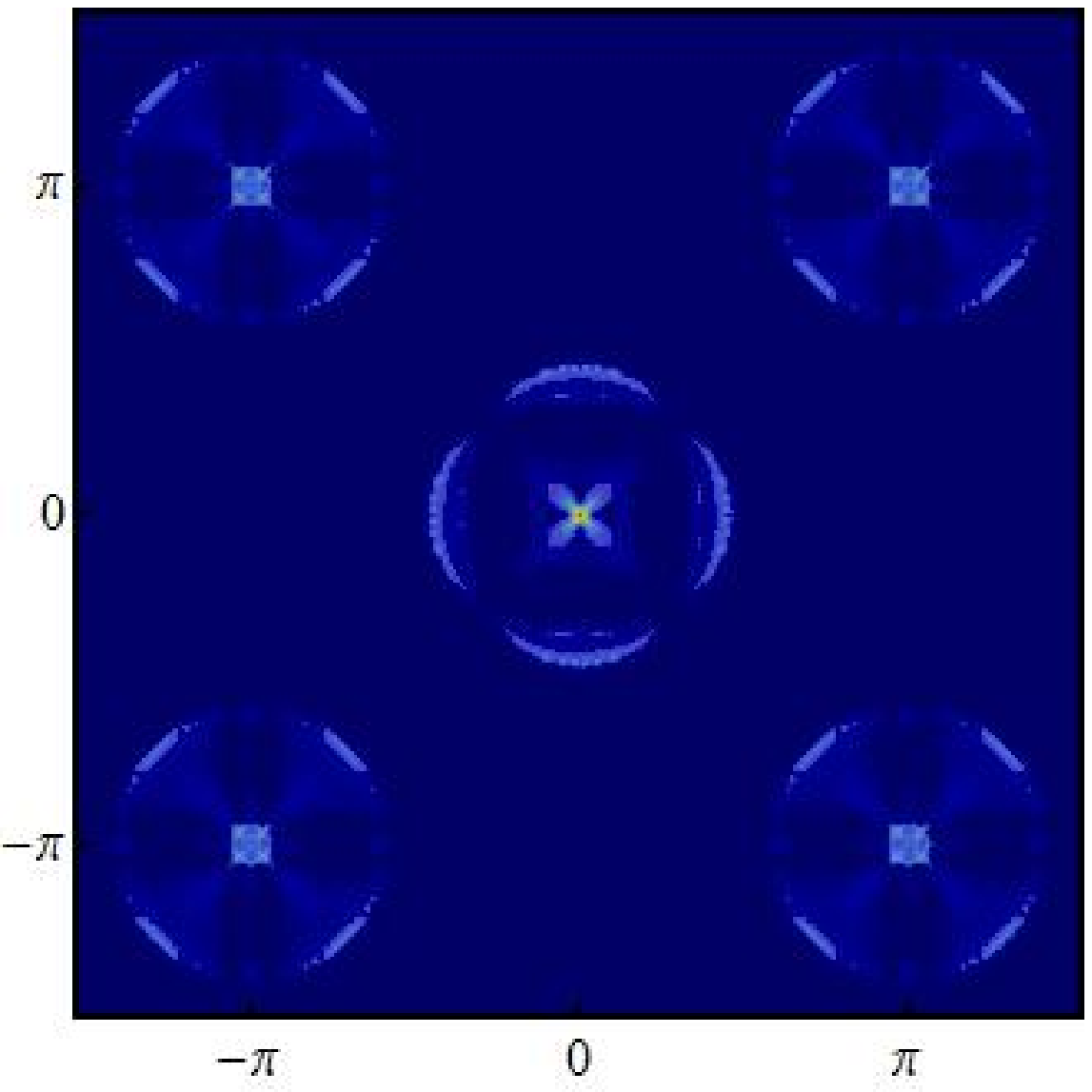} \\
\includegraphics[width=0.18\textwidth]{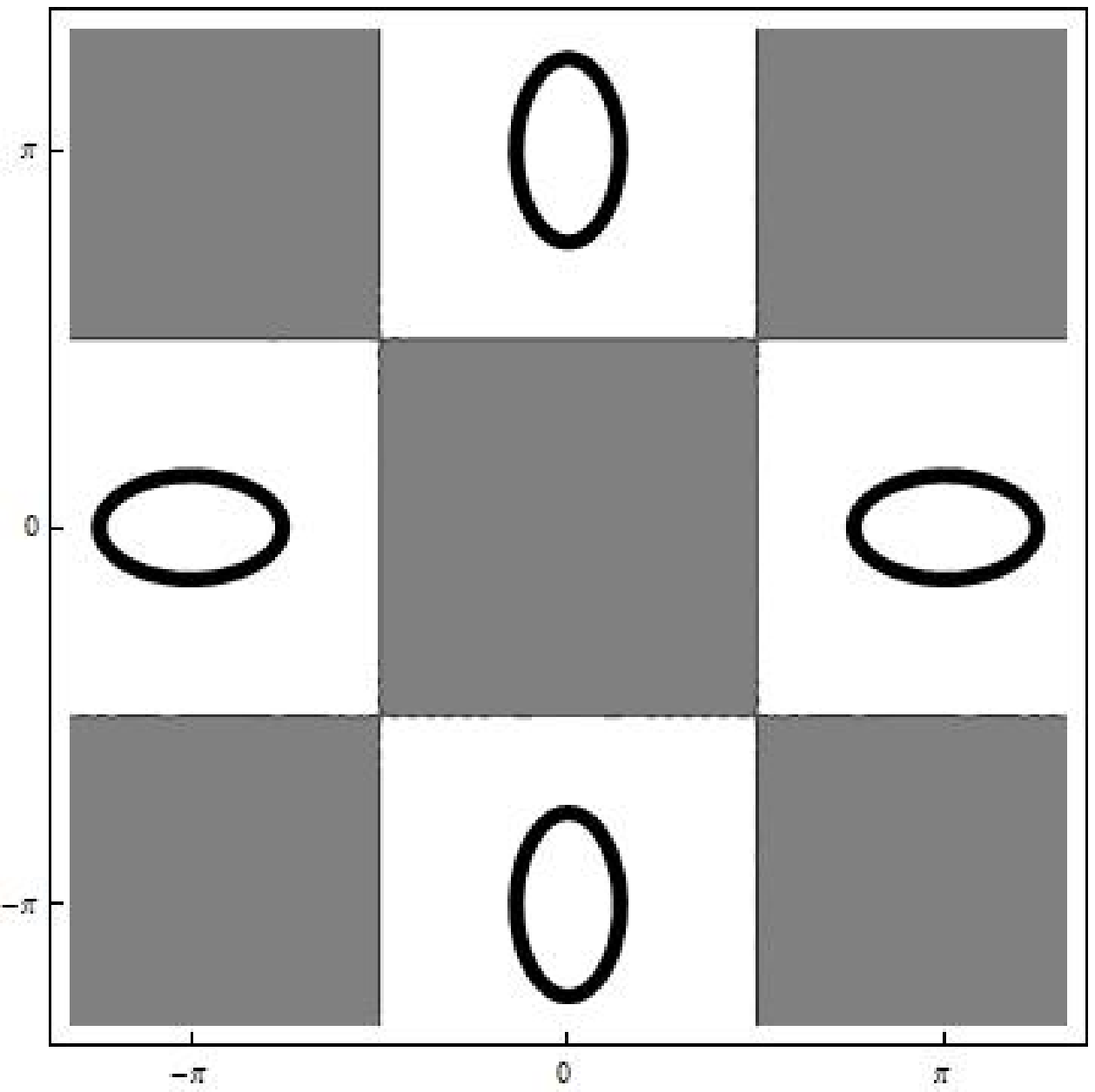} & \includegraphics[width=0.18\textwidth]{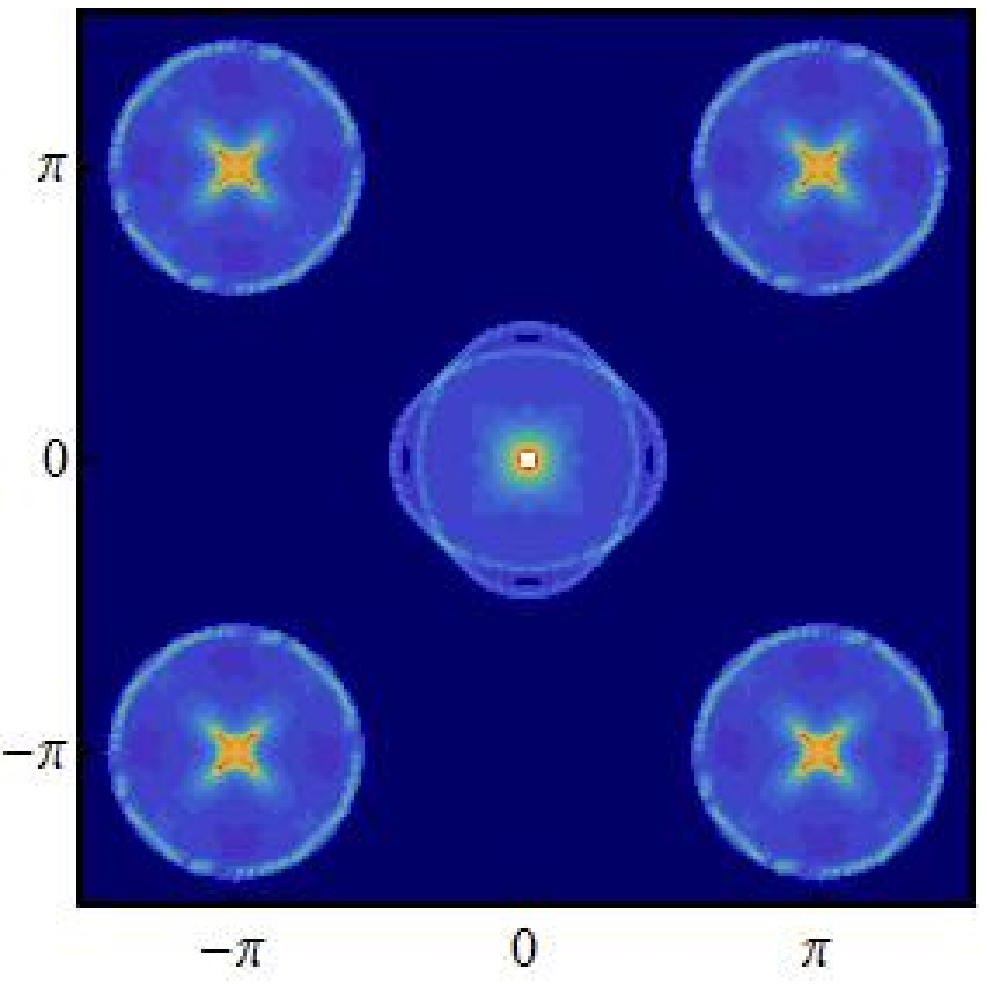} & \includegraphics[width=0.18\textwidth]{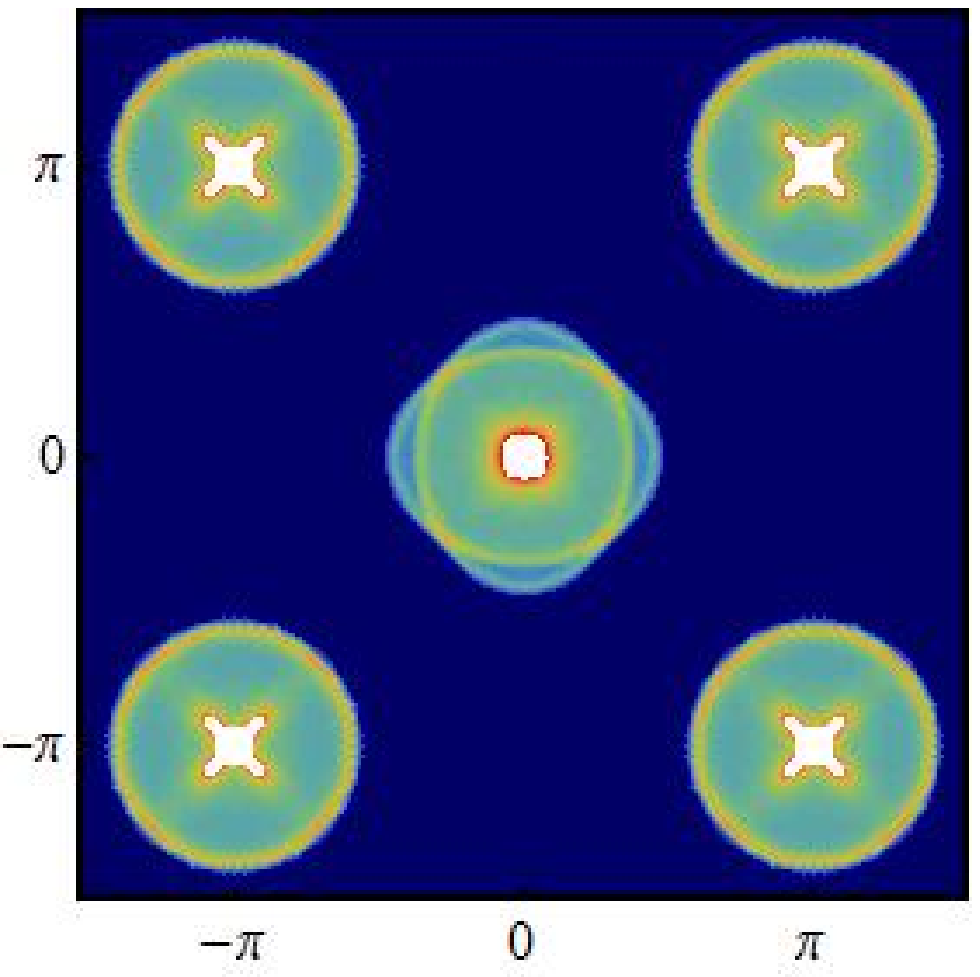} & \includegraphics[width=0.18\textwidth]{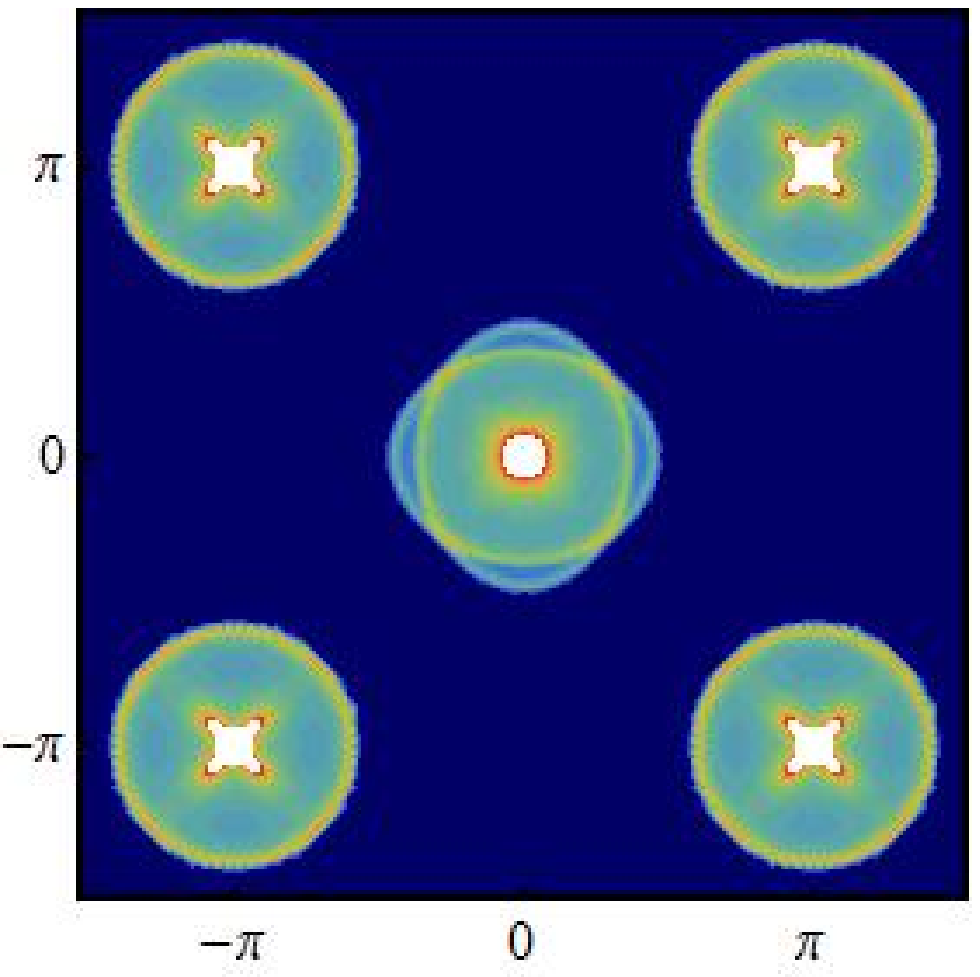}
\end{tabular}
\includegraphics[width=0.25\textwidth,angle=90]{colorfunction.eps}
\caption{(Color online) (a): Fermi surfaces in K$_x$Fe$_{2-y}$Se$_2$. The gray/white color represents positive/negative sign of SC gap with nodeless $d$- (1st row), extended $s$- (2nd row) and $s_{++}$-wave (3rd row) pairing symmeties. The GJDOS pattern for (b) nonmagnetic (c) magnetic and (d) pair-breaking ($i=1$) impurity cases.
Parameters are set as: $t=0.7, \mu=-1.266, \lambda=0.2, \Delta =0.01, \eta =0.005, \omega=0.01$.
}\label{fig3}
\end{figure*}
We then investigate the QPI pattern of the recently discovered K$_{x}$Fe$_{2-y}$Se$_{2}$ superconductor. This iron-based superconductor is interesting because the hole pocket at the center is merged beneath the Fermi level and only the electron pockets are left over. It calls for a re-evaluation of the spin fluctuation pairing mechanism due to the absence of the hole pocket and also raises a question about the possible superconducting pairing symmetry in K$_{x}$Fe$_{2-y}$Se$_{2}$.  Nodeless $d_{x^2-y^2}$-wave~\cite{Seo,feng,Lee,Scalapino2,Tetsuro Saito}, nodal extended $s$-wave~\cite{Lee} and $s_{++}$-wave~\cite{Seo,feng,Tetsuro Saito} pairing symmetries have been suggested. While the last one is just a remainder of $s_\pm$ with the same gap sign on the electron pockets, pairing states of the other two are fascinating for their complex gap structures. For the nodeless $d$-wave pairing, the gap function is slightly anisotropic without sign change around each pocket, while changes its sign between pockets, as schematically illustrated in upper-left panel of Fig.~\ref{fig3}. As for the extended $s$-wave pairing, the gap function around each pocket is similar to the $d$-wave pairing of the cuprates and this gap structure is rotated by $\pi/2$ between pockets, as shown in the middle-left panel of Fig.~\ref{fig3}. Distinct gap structures of different pairing symmetries indicates featured QPI pattern for various types of impurity. In return, the pairing symmetry for K$_{x}$Fe$_{2-y}$Se$_{2}$ can be deduced by checking the QPI patterns from FT-STS experiments.

The normal-state dispersions of K$_{x}$Fe$_{2-y}$Se$_{2}$ are described by
$\xi_\mathbf{k}=-t[(1-\lambda)\cos{(k_x-\pi)}+(1+\lambda)\cos{k_y}]-\mu$
for the electron pocket located around $(\pi,0)$ and $\xi_\mathbf{k}=-t((1+\lambda)\cos{k_x}+(1-\lambda)\cos{(k_y-\pi)})-\mu$
for the electron pocket located around  $(0,\pi)$.
$\lambda$ accounts for the ellipticity of the pockets.
Gap functions are set as: $\Delta (\cos{k_x}-\cos{k_y})$ for the nodeless $d_{x^2-y^2}$-wave, $\Delta (\cos{k_x}+\cos{k_y})$ for the nodal extended $s$-wave, and $\Delta \cos{k_x}\cos{k_y}$ for the $s_{++}$-wave pairing states.
In Fig.~\ref{fig3}, we show the GJDOS patterns for three types of impurity with nodeless $d$-wave, extended $s$-wave and $s_{++}$-wave pairing symmetries. While the three patterns seem to be similar for the nonmagnetic case, they are distinct for the pair-breaking case, especially around nesting vectors $\mathbf{Q}=(\pm\pi,\pm\pi)$. Those changes can be summarized as follows: with stronger pair-breaking scattering (by increasing the magnetic field experimentally), the GJDOS magnitude at $\mathbf{Q}$ is strongly enhanced for $s_{++}$-wave while deeply suppressed for the nodeless $d$-wave and extended $s$-wave pairing states due to the sign difference of SC gaps between the two electron pockets. Furthermore, the nodeless $d$ can be differentiated from the extended $s$ by examining the enhancement or suppression of QPI signal around the center. Thus by introducing more pair-breaking scattering, which can be realized by enhancing the magnetic field, one can deduce the pairing symmetry of K$_{x}$Fe$_{2-y}$Se$_{2}$ by checking the magnetic-field dependence of the QPI patterns.

In summary, we have introduced a new quantity GJDOS and employed it to study QPI in superconductors. It has been shown that the the GJDOS can faithfully reflect both the appropriate coherent factors and the singular behavior of FT-LDOS. For $d$-wave superconductors with the Dirac cone approximation, analytic results have been obtained for both FT-LDOS and GJDOS, finding that they share the same angular factors (arising from the coherent factors) as well as the singular boundaries. This sharing, which may be rather general as we further argued, can account for why GJDOS is appropriate to explain QPI in superconductors. We have also shown that GJDOS 
is a powerful tool in revealing the SC gap structure, and our numeral results agree very well with the STS experiments for $d$-wave cuprates and $s_{\pm}$-wave Fe(Se,Te). We have further used the GJDOS approach to explore the possible pairing symmetry of K$_{x}$Fe$_{2-y}$Se$_{2}$, suggesting that FT-STS experiments may be rather promising for revealing its pairing symmetry. 
Finally, we wish to pinpoint that the GJDOS approach may also be applied to QPI in topological insulator~\cite{Beidenkopf}, where 
the nontrivial spin-orbit coupling
may bring in an interesting effective `coherent factor'.

The authors acknowledge a helpful communication with T.~Hanaguri. The work was supported by the RGC of Hong Kong under Grant
No.HKU7055/09P and HKUST3/CRF/09, the URC fund of HKU, the Natural
Science Foundation of China No.10674179.


\begin{thebibliography}{99}
\bibitem{note1}
It is readily to derive the corresponding formula for $\omega<0$.
\end{thebibliography}
\end{document}